\documentstyle[12pt,epsf]{article}
\newcommand\slv{v\kern-5pt\raise1pt\hbox{$\scriptstyle/$}}
\newcommand\VV{\setbox0=\hbox{V}\hbox{\rm V\raise\ht0
  \hbox to0pt{\hss\vbox to0pt{\hbox{v}\vss}}}}

\begin{document}

\thispagestyle{empty}
\begin{flushright}
MZ-TH/96-06\\
BudkerINP/96-21\\
April 1996\\
\end{flushright}
\vspace{0.5cm}
\begin{center}
{\Large\bf Two-loop Anomalous Dimensions}\\[.3cm]
{\Large\bf of Heavy Baryon Currents in}\\[.3cm]
{\Large\bf Heavy Quark Effective Theory}\\
\vspace{1.3cm}
S. Groote,\footnote{Groote@dipmza.physik.uni-mainz.de}
J.G. K\"orner\footnote{Koerner@dipmza.physik.uni-mainz.de}
and O.I. Yakovlev\footnote{Yakovlev@dipmza.physik.uni-mainz.de}
\footnote{Permanent address:
Budker Institute for Nuclear Physics, 630090, Novosibirsk, Russia}
\\[1cm]
Institut f\"ur Physik, Johannes-Gutenberg-Universit\"at,\\[.2cm]
Staudinger Weg 7, D-55099 Mainz, Germany\\
\end{center}
\vspace{1cm}
\begin{abstract}\noindent 
We present results on the two-loop anomalous dimensions of the heavy baryon 
HQET currents $J=(q^TC\Gamma\tau q)\Gamma'Q$ with arbitrary Dirac matrices 
$\Gamma$ and $\Gamma'$. From our general result we obtain the two-loop 
anomalous dimensions for currents with quantum numbers of the ground state 
heavy baryons $\Lambda_Q$, $\Sigma_Q$ and $\Sigma_Q^*$. As a by-product of 
our calculation and as an additional check we rederive the known two-loop 
anomalous dimensions of mesonic scalar, pseudoscalar, vector, axial vector 
and tensor currents $(J=\bar q\Gamma q)$ in massless QCD as well as in HQET. 
\end{abstract}


\newpage

\section{Introduction}
The investigation of the renormalization-group (RG) properties of composite 
operators in QCD is an important ingredient in the numerous applications of 
QCD in particle physics. In the present paper we shall consider some aspects 
of the RG properties of heavy baryonic currents in the framework of the 
Heavy Quark Effective Theory (HQET)~\cite{Eich,Grin}, in particular we 
consider the anomalous dimensions of the currents associated with the 
ground-state heavy baryons. It is well known that the heavy mass dependence 
of operators and matrix elements in QCD can be very well organized by 
employing an $1/m_Q$-expansion provided by HQET. Here we work to lowest 
order in the heavy mass expansion, i.e.\ we treat the heavy quark fields as 
effective static fields. While this approximation should be a good starting 
point to understand the properties of heavy hadrons containing one heavy 
quark ($b$ or $c$) it leads to significant simplifications in the 
calculation of the Wilson coefficients that appear in the Operator Product 
Expansion (OPE) and in the calculation of RG coefficients such as the 
anomalous dimension coefficient dealt with in this paper.

The anomalous dimensions of composite operators are basic ingredients in 
all applications of the OPE method in QCD such as QCD sum rules~\cite{Vain} 
and related approaches. In this paper we undertake to calculate the two-loop 
anomalous dimensions of heavy baryon currents in the static approximation. 
The anomalous dimensions of the baryonic currents are one of the ingredients 
needed in the analysis of the correlators of two baryonic currents via their 
short distance expansion. The ensuing QCD sum rules allow one to compute the 
mass of heavy baryons and their residues in terms of basic non-perturbative 
QCD parameters. 

Another important application where the anomalous dimensions enter is the 
determination of the universal non-perturbative baryonic Isgur-Wise function 
$\xi(\omega)$~\cite{Isgur} from three-point QCD sum rules. Of particular 
importance is the calculation of the slope of the Isgur-Wise function at 
zero recoil. Knowledge of the two-loop anomalous dimensions is important 
when one discusses the matching of HQET baryonic currents with the 
corresponding currents in full QCD. Furthermore, the anomalous dimensions are 
needed for the calculation of some basic QCD matrix elements that appear in 
QCD sum rules, as for example the expectation value of the kinetic energy 
$\mu_\pi^2$, the expectation value of the chromomagnetic energy $\mu^2_G$, 
the spin-dependent axial current matrix element $\mu_s^2$, the couplings 
of heavy baryons with pions and photons, and the magnetic and axial moments 
of heavy baryons.

The general structure of heavy baryon currents has the form (see 
e.g.~\cite{GrYa} and refs.\ therein)
\begin{equation}\label{current}
J=[q^{iT}C\Gamma\tau q^j]\Gamma'Q^k\epsilon_{ijk}.
\end{equation}
Here the index $T$ means transposition, $C$ is the charge conjugation 
matrix with the properties $C\gamma^T_\mu C^{-1}=-\gamma_\mu$ and 
$C\gamma^T_5C^{-1}=\gamma_5$, $i,j,k$ are colour indices and $\tau$ is a 
matrix in flavour space. The effective static field of the heavy quark is 
denoted by~$Q$.

The currents of the $\Lambda_Q$ and the heavy quark spin doublet 
$\{\Sigma_Q,\Sigma_Q^*\}$ are associated with the spin-parity quantum 
numbers $j^P=0^+$ and $j^P=1^+$ for the light diquark system, respectively. 
For each of the ground state baryon currents there are two independent 
current components $J_1$ and $J_2$. They are given by~\cite{Shuryak,GrYa}
\begin{eqnarray}
J_{\Lambda1}&=&[q^{iT}C\tau\gamma_5q^j]Q^k\varepsilon_{ijk},\qquad
J_{\Lambda2}\ =\ [q^{iT}C\tau\gamma_5\gamma_0q^j]Q^k\varepsilon_{ijk},
  \nonumber\\[7pt]
J_{\Sigma1}&=&[q^{iT}C\tau\vec\gamma q^j]
  \cdot\vec\gamma\gamma_5Q^k\varepsilon_{ijk},\qquad
J_{\Sigma2}\ =\ [q^{iT}C\tau\gamma_0\vec\gamma q^j]
  \cdot\vec\gamma\gamma_5Q^k\epsilon_{ijk},\nonumber\\
\vec J_{\Sigma^*1}&=&[q^{iT}C\tau\vec\gamma q^j]Q^k\varepsilon_{ijk}
  +\frac13\vec\gamma[q^{iT}C\tau\vec\gamma q^j]
  \cdot\vec\gamma Q^k\varepsilon_{ijk},\label{currents}\\
\vec J_{\Sigma^*2}&=&[q^{iT}C\tau\gamma_0\vec\gamma q^j]Q^k\varepsilon_{ijk}
  +\frac13\vec\gamma[q^{iT}C\gamma_0\vec\gamma q^j]
  \cdot\vec\gamma Q^k\varepsilon_{ijk},\nonumber
\end{eqnarray}
where $\vec J_{\Sigma^*1}$ and $\vec J_{\Sigma^*2}$ satisfy the spin-$3/2$ 
condition $\vec\gamma\vec J_{\Sigma^*i}=0$ ($i=1,2$). The flavour matrix 
$\tau$ is antisymmetric for $\Lambda_Q$ and  symmetric for the heavy quark 
spin doublet $\{\Sigma_Q,\Sigma_Q^*\}$. The currents written down in 
Eq.~(\ref{currents}) are rest frame currents. The corresponding expressions 
in a general frame moving with velocity four-vector $v^\mu$ can be obtained 
by the substitutions $\gamma_0\rightarrow\slv$ and 
$\vec\gamma\rightarrow\gamma^\mu_\perp=\gamma^\mu-\slv v^\mu$. 

The one-loop renormalization of these currents was considered in~\cite{GrYa}.
The renormalization of the corresponding light baryonic currents to one-loop 
order was presented in~\cite{Peskin} and to two-loop order in~\cite{PivSur}. 
In this paper we determine the two-loop anomalous dimensions of the heavy 
baryon currents. We evaluate the numerous two-loop diagrams with the help of 
REDUCE and MATHEMATICA packages written by us, which are based on the 
recurrence-relations in QCD and HQET presented in~\cite{ChTk,SuTk,BrGr,David}.

The paper is organized as follows. In Sec.~2 we furnish some general 
remarks on anomalous dimensions, discuss the one-loop renormalization of 
the general heavy baryon currents~(\ref{current}) and give results on the 
one-loop anomalous dimensions in a general covariant gauge. In Sec.~3 
we discuss general features of the two-loop renormalization procedure and 
present our general two-loop results, where we specify to the Feynman gauge. 
In Sec.~3 we also derive a powerful consistency check on our calculation, 
and we compare our results with previously published two-loop results in 
the mesonic sector. In Sec.~4 we consider applications of our general 
results to the ground-state heavy baryon currents~(\ref{currents}). We work 
in the $\overline{\rm MS}$-renormalization scheme throughout. As concerns 
the treatment of $\gamma_5$ we give our results for two different schemes, 
namely the 't~Hooft-Veltman $\gamma_5$-scheme and the naive anticommuting 
$\gamma_5$-scheme. Finally, we show how the results in the two 
$\gamma_5$-schemes become related through a finite renormalization 
prescription as has been demonstrated previously in a different context by 
Trueman and Larin~\cite{True,Lari}. Sec.~5 contains our conclusions.

\newpage

\section{Renormalization procedure and one loop result}

The renormalization procedure consists in the renormalization of all bare 
UV-diver\-gent operators $O_0$ in terms of a renormalization factor $Z$ 
according to
\begin{equation}
O_0=ZO.
\end{equation}
Using dimensional regularization in $D=4-2\epsilon$ dimensions and the 
$\overline{\rm MS}$-scheme~\cite{tHVt}, the factor $Z$ is a polynomial in 
inverse powers of $\epsilon$. The $Z$-factor is constructed in such a way 
that the operator $O$ is finite in physical four-dimensional space. After 
renormalization, the operator $O$ can be seen to depend on the subtraction 
scale (normalization point) $\mu$. The anomalous dimension of the operator 
$O$ is then defined by
\begin{equation}
\gamma=\frac{d\ln Z(\alpha(\mu),a;\epsilon)}{d\ln (\mu)},
\end{equation}
where $a$ is the renormalized gauge parameter in the general covariant 
gauge (with a gluon propagator proportional to 
$-g_{\mu\nu}+(1-a)k_\mu k_\nu/k^2$) and $\alpha(\mu)$ is the renormalized 
coupling constant in four-dimensional space. The $Z$-factor and thereby the 
anomalous dimension $\gamma$ is of course specific to the operator which it 
renormalizes. Thus they would carry an additional index which we suppress 
for the moment. The unrenormalized gauge parameter $a_0$ and the 
unrenormalized coupling constant $\alpha_0$ defined in $D$ dimensions are 
related to the corresponding renormalized quantities $a$ and $\alpha(\mu)$ 
in four dimensions by 
\begin{eqnarray}\label{conn}
\alpha_0=\alpha(\mu)\mu^{2\epsilon}Z_\alpha(\alpha(\mu),a;\epsilon),\qquad
a_0=aZ_3(\alpha(\mu),a;\epsilon).
\end{eqnarray}
One-loop $\overline{\rm MS}$-results for the factors $Z_\alpha$ and $Z_3$ 
have been given e.g.\ in~\cite{PaTa}. They read
\begin{eqnarray}
Z_\alpha&=&1-\frac{\alpha_s}{4\pi\epsilon}
\left[\frac{11}3C_A-\frac43T_FN_F\right],\label{zal}\\ 
Z_3&=&1+\frac{\alpha_s}{4\pi\epsilon}
  \left[\frac{13-3a}6C_A-\frac43T_FN_F\right].\label{z3}
\end{eqnarray}
Note that there is a sign difference in the second $O(\alpha_s)$ terms 
in Eqs.~(\ref{zal}) and~(\ref{z3}) relative to that in~\cite{PaTa} due to 
a difference in the sign convention for $\epsilon$. The factors $Z_\alpha$ 
and $Z_3$ determine two universal RG functions $\beta$ and $\delta$ which 
denote the anomalous dimensions of the coupling and gauge parameters, i.e.
\begin{equation}\label{betadelta}
\beta(\alpha(\mu),a)=-\frac{d\ln Z_\alpha(\alpha(\mu),a;\epsilon)}{d\ln(\mu)}
  \quad\mbox{and}\quad
\delta(\alpha(\mu),a)=-\frac{d\ln Z_3(\alpha(\mu),a;\epsilon)}{d\ln(\mu)}.
\end{equation}
In perturbation theory the renormalization factor or function 
$Z(\alpha(\mu),a;\epsilon)$ will be a double series in the renormalized 
coupling constant $\alpha_s=\alpha(\mu)$ and in inverse powers of $\epsilon$. 
We therefore write the double series expansion
\begin{equation}
Z=\sum_{m=1}^\infty\sum_{k=1}^m\left(\frac{\alpha_s}{4\pi}\right)^m
  \frac1{\epsilon^k}Z_{m,k}=\sum_{k=1}^\infty\frac1{\epsilon^k}Z_k.
\end{equation}
In contradistinction to this the anomalous dimension $\gamma$, as well as the 
functions $\beta$ and $\delta$, are only expanded in powers of the coupling 
constant $\alpha_s$ since they are finite quantities for $D\rightarrow 4$. 
One has
\begin{equation}\label{series}
\gamma=\sum_{m=1}^\infty\left(\frac{\alpha_s}{4\pi}\right)^m\gamma^{(m)},\qquad
\beta=\sum_{m=1}^\infty\left(\frac{\alpha_s}{4\pi}\right)^m\beta^{(m)},\qquad
\delta=\sum_{m=1}^\infty\left(\frac{\alpha_s}{4\pi}\right)^m\delta^{(m)}.
\end{equation}
Using these functions and differentiating $Z(\alpha(\mu),a;\epsilon)$ with 
respect to $\ln(\mu)$ one can compare coefficients at different orders of 
$1/\epsilon^k$~\cite{PaTa}. For $k=0$ one obtains
\begin{equation}\label{anom}
\gamma=-2\frac{\partial Z_{1}}{\partial\ln\alpha_s}
\end{equation}
and for $k>0$
\begin{equation}\label{consistency}
-2\frac{\partial Z_{k+1}}{\partial\ln\alpha_s}
  =\left(\gamma-\beta\frac\partial{\partial\ln\alpha_s}
  -\delta\frac\partial{\partial\ln a}\right)Z_{k}.
\end{equation}
The relations (\ref{anom}) and~(\ref{consistency}) allow one to determine 
the anomalous dimension on the one hand and, on the other hand, provide for 
a connection between the coefficients $Z_{m,k}$ at different powers of 
$1/\epsilon$ which allows one to check on the consistency of the two-loop 
calculations. For example, the first order contributions in $1/\epsilon$ 
determine the one- and two-loop anomalous dimensions
\begin{equation}\label{andim12}
\gamma^{(1)}=-2Z_{1,1}\quad\mbox{and}\quad\gamma^{(2)}=-4Z_{2,1}.
\end{equation}
For the two-loop coefficients, Eq.~(\ref{consistency}) leads to the 
consistency condition
\begin{equation}\label{check}
-4Z_{2,2}=\left(\gamma^{(1)}-\beta^{(1)}-\delta^{(1)} 
\frac\partial{\partial\ln a}\right)Z_{1,1}.
\end{equation}

\subsection{One-loop results}

Let us first consider the one-loop renormalization of the effective heavy 
baryon currents~(\ref{current}). The bare light quark and bare effective 
heavy quark fields are related to the renormalized fields by
\begin{equation}\label{fieldrenorm}
q_0=Z_q^{1/2}q,\qquad Q_0=Z_Q^{1/2}Q.
\end{equation}
In the $\overline{\rm MS}$-scheme with $D=4-2\epsilon$ space-dimensions we 
have 
\begin{equation}\label{ZqQ}
Z_q=1-a_0\frac{g_0^2C_F}{(4\pi)^2\epsilon},\qquad
Z_Q=1+(3-a_0)\frac{g_0^2C_F}{(4\pi)^2\epsilon},
\end{equation}
where $g_0$ is the bare coupling in QCD and $a_0$ is the bare gauge 
parameter. We use the usual definitions for $SU(N)$, i.e.\ 
$C_F=(N_c^2-1)/2N_c$, $C_A=N_c$, $C_B=(N_c+1)/2N_c$, and $T_F=1/2$ for 
$N_c=3$, $N_F$ being the number of light quarks. The bare current is 
renormalized by the factor $Z_J$, i.e.
\begin{equation}\label{renormj}
J_0=(q_0^TC\Gamma\tau q_0)\Gamma'Q_0=Z_qZ^{1/2}_QZ_VJ=Z_JJ.
\end{equation}
As can be seen from Eq.~(\ref{renormj}), the anomalous dimension of the full 
current $J$ is composed of three parts associated with the renormalization 
of the light and heavy quark fields, and the renormalization of the vertex. 
One thus has
\begin{equation}\label{gammaj}
\gamma_J=2\gamma_q+\gamma_Q+\gamma_V.
\end{equation}
The one-loop anomalous dimensions $\gamma_q^{(1)}$ and $\gamma_Q^{(1)}$ can 
be read off Eqs.~(\ref{fieldrenorm}) and~(\ref{ZqQ}) and are
\begin{equation}\label{gamma1legs}
\gamma_q^{(1)}=C_Fa,\qquad\gamma_Q^{(1)}=C_F(a-3).
\end{equation}
A concerns the corrections to the vertex $(q_0^TC\Gamma\tau q_0)\Gamma'Q_0$, 
one obtains
\begin{equation}\label{zgam1}
Z_V=1+\frac{\alpha_sC_B}{4\pi\epsilon}((n-2)^2+3a-1).
\end{equation}
The one-loop vertex $Z$-factor has been calculated in the general covariant 
gauge \hbox{$a\ne 1$} since the consistency condition in Eq.~(\ref{check}) 
requires knowledge of the dependence of $Z_{1,1}$ on the gauge parameter $a$. 
In deriving Eq.~(\ref{zgam1}) we have parameterized the general light-side 
Dirac structure $\Gamma$ in terms of antisymmetrized products of $n$ Dirac 
matrices, $\Gamma=\gamma^{[\mu_1}\cdots\gamma^{\mu_n]}$. For the time being 
we assume that there is no $\gamma_5$ in the Dirac string $\Gamma$ and 
postpone the discussion on how to define $\gamma_5$ in $D\ne 4$ dimensions 
to the end of the paper when we present explicit values for the anomalous 
dimensions. The above form for the Dirac string $\Gamma$ introduces a $n$- 
and $s$-dependence in the vertex corrections due to the identities
\begin{equation}\label{Dirac}
\gamma_\alpha\Gamma\gamma_\alpha=h\Gamma=(-1)^n(D-2n)\Gamma,\qquad
  \gamma_0\Gamma\gamma_0=(-1)^ns\Gamma.
\end{equation}
Using the results of Eq.~(\ref{andim12}) we immediately obtain the one-loop 
anomalous dimensions of the irreducible vertex
\begin{equation}\label{an1}
\gamma^{(1)}_V=-2C_B((n-2)^2+3a-1).
\end{equation}
According to Eqs.~(\ref{gammaj}) and~(\ref{gamma1legs}), the one-loop 
anomalous dimension of the baryonic current is given by
\begin{equation}\label{an11}
\gamma_J=\frac{\alpha_s}{4\pi}\Big(-2C_B((n-2)^2+3a-1)+3C_F(a-1)\Big)
  +O(\alpha_s^2).
\end{equation}
Note that the dependence on the gauge parameter $a$ drops out of 
Eq.~(\ref{an11}) when substituting $C_F=4/3$ and $C_B=2/3$. This is a 
necessary requirement since the anomalous dimension is a gauge-independent 
quantity. When one specifies to the Feynman gauge ($a=1$), the results in 
Eqs.~(\ref{zgam1}), (\ref{an1}) and~(\ref{an11}) agree with the 
corresponding results presented in~\cite{GrYa}.

\section{Two-loop renormalization}

In this section we consider the two-loop renormalization of the heavy 
baryon current~(\ref{current}) in the $\overline{\rm MS}$-scheme where we 
now work in the Feynman gauge to simplify the calculation. First of all, 
the two-loop anomalous dimensions of the quark fields are well known and 
can be taken e.g. from~\cite{BrGr,Tara,Jone,Gime,JiMu}. 
They are given by
\begin{equation}\label{gammaq}
\gamma^{(2)}_q=C_F\left(\frac{17}2C_A-2T_FN_F-\frac32C_F\right),\qquad
\gamma^{(2)}_Q=C_F\left(-\frac{38}3C_A+\frac{16}3T_FN_F\right).
\end{equation}
In order to determine the vertex renormalization factor $Z_V$ that 
renormalizes the bare proper vertex according to $\VV_0=Z_V\VV$ we need to 
compute the bare proper vertex $\VV_0$ in one-loop and two-loop order. The 
Dirac structure of the unrenormalized and renormalized vertices $\VV_0$ and 
$\VV$ is given by the Born-term Dirac structure of the baryonic currents, 
i.e.\ by $\Gamma\otimes\Gamma'$ in the notation of Eq.~(\ref{current}) (all 
heavy baryon currents are renormalized multiplicatively in the effective 
theory!). We can thus write $\VV_0=V_0\Gamma\otimes\Gamma'$ and 
$\VV=V\Gamma\otimes\Gamma'$, where $V$ is finite. By requiring $V=V_0/Z_V$ 
to be finite one can calculate $Z_V$ in terms of $V_0$. For this purpose one 
first evaluates the one- and two-loop diagrams in terms of the bare coupling 
constant $g_0$ and, in case of the one-loop diagrams, in terms of the bare 
gauge parameter $a_0$. One then substitutes for these in terms of the 
renormalized $g_s$ ($g_s^2=4\pi\alpha_s$) and $a$ with the help of the 
Eqs.~(\ref{conn}), (\ref{zal}) and~(\ref{z3}). The resulting expression for 
the bare proper vertex $V_0$ may again be presented as a double series in 
$\alpha_s$ and $\epsilon$,
\begin{equation}\label{vseries}
V_0=\sum_{m=0}^\infty\sum_{k=0}^m\left(\frac{\alpha_s}{4\pi}\right)^m
  \frac1{\epsilon^k}V_{m,k}.
\end{equation}
Thus the coefficients $V_{m,k}$ and $Z_{m,k}$ become related. For example, 
for the first few coefficients one has the relations
\begin{equation}\label{zvrel}
Z_{1,1}=V_{1,1},\qquad Z_{2,2}=V_{2,2},\qquad
  Z_{2,1}=V_{2,1}-V_{1,1}V_{1,0}.
\end{equation}
Note that in the relation for the $Z$-factor $Z_{2,1}$ one also 
has to include one-loop contributions.

Three different contributions to the two-loop vertex corrections $V_0$ may 
be identified. First one has the set of two-loop graphs where the two-loop 
contributions are associated with one of the heavy-light subsystems. Then 
there is the subset of two-loop graphs that are associated with the 
light-light system. Finally there are the irreducible contributions where 
the two loops connect all three quark lines. Hence we write
\begin{equation}
V_0=2V^{(hl)}_0+V^{(ll)}_0+V^{(ir)}_0,
\end{equation}
where the factor of two on the r.h.s.\ accounts for the fact that there are 
two possible heavy-light configurations. The three types of contributions 
will be considered in turn in the following subsections.

\subsection{The heavy-light system}

In this subsection we consider composite operators $(qQ)$ with one massless 
quark field $q$ and one effective static heavy quark field $Q$. There are 
11 different two-loop diagrams contributing to the composite operator $(qQ)$ 
which are shown in Fig.~1. We choose the momentum of the light quark to be 
zero and regularize all IR-divergencies by taking the heavy quark off its 
mass-shell. The bare proper vertex $V^{(hl)}_0$ of the heavy-light system 
is calculated to two-loop order in the Feynman gauge ($a=1$) using the 
algorithm developed in~\cite{BrGr}. The result including the Born-term and 
one-loop contributions can be presented as
\begin{equation}\label{verthl}
V^{(hl)}_0=1+\lambda_IC_0b^{(hl)}_0
  +\sum_{i=1}^{11}\lambda_I^2C_ib^{(hl)}_i,
\end{equation}
where we use the notation $\lambda_I:=(g_0^2/(4\pi)^{D/2})(-2\omega)^{D-4}$. 
The contributions of the various diagrams in Fig.~1 are separately 
identified according to $i=0$ for the one-loop contribution and 
$i=1,\ldots,11$ for the two-loop contributions. The associated colour 
factors are denoted by $C_i$,
\begin{eqnarray}
C_0=C_B,\qquad\qquad
C_1=C_B^2,\qquad\qquad
C_2=C_B(C_B-{\textstyle\frac12}C_A),\nonumber\\[7pt]
C_3=C_4=C_BC_F,\qquad\qquad\qquad
C_5=C_6=C_B(C_F-{\textstyle\frac12}C_A),\label{colourfac}\\[7pt]
C_7=C_BN_FT_F,\quad
C_8=C_{11}=C_BC_A,\quad
C_9=C_{10}={\textstyle\frac12}C_BC_A.\nonumber
\end{eqnarray}

One should keep in mind that the colour factors have to be calculated for 
the completely antisymmetric baryonic colour configuration 
$q^iq^jQ^k\varepsilon_{ijk}$. Because of the difference in colour structure 
it is not a priori clear how the baryonic heavy-light two-loop 
results can be related to the mesonic case with colour structure 
$\bar q_iQ^j\delta^i_j$. However, it is not difficult to see that the 
corresponding colour-singlet colour factors for the mesonic case can be 
obtained from Eq.~(\ref{colourfac}) by the substitution $C_B\rightarrow C_F$. 
Note, though, that this argument cannot be turned around. The contributions 
to the baryonic anomalous dimensions cannot be read off from the mesonic ones.

The coefficients $b^{(hl)}_i$ are listed in Appendix~A. The coefficients 
$V^{(hl)}_{n,k}$ are then given by
\begin{eqnarray}
V^{(hl)}_{1,1}\ =\ C_Ba,\qquad V^{(hl)}_{1,0}&=&0,\\
V^{(hl)}_{2,2}\ =\ C_B\Big(\frac12C_B-C_A\Big),\qquad
V^{(hl)}_{2,1}&=&-C_B\Big(C_B(1-4\zeta(2))-C_A(1-\zeta(2))\Big).\nonumber
\end{eqnarray}
Using the relations~(\ref{zvrel}) one can then calculate the coefficients 
$Z_{n,k}$ which determine the two-loop anomalous dimension of the 
subset of the heavy-light graphs (with antisymmetric colour structure),
\begin{equation}
\gamma^{(2)}_{(hl)}=C_B^2(4-16\zeta(2))-C_BC_A(4-4\zeta(2)).
\label{anomhl}
\end{equation}
There are two checks on our calculation. First, the coefficients $Z_{2,2}$ 
and $Z_{1,1}$ can be seen to satisfy the consistency condition 
Eq.~(\ref {check}). Second, as remarked on above, it is not difficult to 
transcribe the calculation to the mesonic case (colour-singlet heavy-light 
quark-antiquark system $\bar q_iQ^j\delta^i_j$ with colour factors given 
after Eq.~(\ref{colourfac})). After transcription to the mesonic case we 
find agreement of our results with the results given in~\cite{BrGr,Gime,JiMu}.

\subsection{The light-light system}

The most difficult part of the calculation is the evaluation of the two-loop 
bare proper vertex of the light-light system. There are again 11 different 
Feynman diagrams corresponding to the two-loop corrections of the vertex 
with two light quarks (see Fig.~1 with the heavy quark line substituted by 
a light quark line).

Again we need to compute the bare proper vertex $V_0$ at one- and two-loop 
order. The external momentum of one of the light quark legs is set to zero. 
Using the power counting method, one can convince oneself that this procedure 
does not introduce any unwanted infrared singularities. All infrared 
singularities are regularized by choosing a nonzero momentum for the second 
light quark leg. On the other hand, setting one external quark momentum to 
zero, the ultraviolet properties of the diagrams are not affected. The 
advantage is that all diagrams now become two-point functions, which can be 
treated by using the algorithm presented in~\cite{ChTk}. In the first stage 
the one- and two-loop integrals are evaluated independently of the specific 
Dirac structure of the Dirac string $\Gamma$. In the second stage one 
specifies $\Gamma$ and the general result can then be written down using the 
identities~(\ref{Dirac}). The results of the calculation are again presented 
in coefficient form according to
\begin{equation}\label{vertll}
V^{(ll)}_0=1+\sum_{j=0}^4\left(\lambda_GC_0b^{(ll)}_{0,j}
  +\sum_{i=1}^{11}\lambda_G^2C_ib^{(ll)}_{i,j}\right )(s(D-2n))^j,
\end{equation}
where $\lambda_G:=(g_0^2/(4\pi)^{D/2})(-1/p^2)^{D/2-2}$. The colour factors 
$C_i$ have been calculated in Eq.~(\ref{colourfac}), the index $i$ is 
defined as in Eq.~(\ref{verthl}).

We need to explain the index $j$ and the factor $(s(D-2n))^j$ appearing in 
Eq.~(\ref{vertll}). In the calculation of the one- and two-loop 
contributions one encounters different contracted forms of the general 
light Dirac string structure $\Gamma$ according to
\begin{eqnarray}
\Gamma_0\ =\ \Gamma,\qquad
\Gamma_1&=&\gamma_\mu\gamma_0\Gamma\gamma_0\gamma_\mu,\qquad
\Gamma_3\ =\ \gamma_\mu\gamma_\nu\gamma_\rho\gamma_0\Gamma
  \gamma_0\gamma_\rho\gamma_\nu\gamma_\mu,\label{contractions}\\
\Gamma_2&=&\gamma_\mu\gamma_\nu\Gamma\gamma_\nu\gamma_\mu,\qquad
\Gamma_4\ =\ \gamma_\mu\gamma_\nu\gamma_\rho\gamma_\sigma\Gamma
  \gamma_\sigma\gamma_\rho\gamma_\nu\gamma_\mu.\nonumber
\end{eqnarray}
Again, we defer the discussion on $\gamma_5$ to Sec.~4 assuming $\Gamma$ to 
contain no $\gamma_5$. From the identities in Eq.~(\ref{Dirac}) one obtains 
$\Gamma_j=(s(D-2n))^j\Gamma$ leading to the factor $(s(D-2n))^j$ in 
Eq.~(\ref{vertll}).

The coefficients $b^{(ll)}_{i,j}$ are listed in Appendix~B. Then, organizing 
the contributions according to the double expansion in Eq.~(\ref{vseries}) 
one has
\begin{eqnarray}
V^{(ll)}_{1,1}&=&C_B\Big((n-2)^2+a-1\Big),\\
V^{(ll)}_{1,0}&=&C_B\Big((n-2)^2+(n-2)(s+2)+a-1\Big),\\
V^{(ll)}_{2,1}&=&\frac{C_B}{36}\Big(
  9C_B(13n^4-96n^3+248n^2-256n+88+4(n-2)^3s)\nonumber\\&&
  -72C_F(n-3)(n-1)-C_A(18n^2-144n^3+289n^2+128n-444)\nonumber\\&&
  -4N_FT_F(n^2-16n+24)\Big),\\
V^{(ll)}_{2,2}&=&\frac{C_B}6\Big(
  3C_B(n-2)^4-C_A(11n^2-44n+39)+4N_FT_F(n-3)(n-1)\Big).\qquad
\end{eqnarray}
As in Sec.~3.1, the one-loop results have been calculated in the general 
covariant gauge whereas the two-loop results are given only for the Feynman 
gauge. The result for the two-loop anomalous dimension of the proper 
light-light vertex is then given by
\begin{eqnarray}
\gamma^{(2)}_{(ll)}&=&\frac{C_B}9\Big(
  -9C_B(5n^4-40n^3+104n^2-96n+24+4(n-2)^3s)\nonumber\\&&
  +72C_F(n-3)(n-1)+C_A(18n^2-144n^3+289n^2+128n-444)\nonumber\\&&
  +4N_FT_F(n^2-16n+24)\Big).
\end{eqnarray}
We have checked that the coefficients $Z^{(ll)}_{2,2}$ and $Z^{(ll)}_{1,1}$ 
satisfy the consistency condition Eq.~(\ref{check}) for any value of $n$. 

A further check of our calculation is provided by the colour singlet 
projection of the proper vertex for the light-light mesonic configuration 
$(\bar q^iq_j\delta_i^j)$ which can be obtained from our result with the 
substitutions $q^TC\rightarrow\bar q$ and $C_B\rightarrow C_F$. 
Collecting the $C_F$-contributions one has
\begin{eqnarray}\label{gam2llmes}
\gamma^{(2)}_{(ll)}&=&\frac{C_F}9\Big(-9C_Fn(n-4)(5n^2-20n+16)\nonumber\\&&
  +C_A(18n^4-144n^3+289n^2+128n-444)\\&& +4N_FT_F(n^2-16n+24)\Big).\nonumber
\end{eqnarray}
After appending the anomalous dimensions $2\gamma_q^{(2)}$ of the light 
quark fields one can then compare with previously published results in the 
mesonic sector. For $n=0$\break Eq.~(\ref{gam2llmes}) coincides with the 
known anomalous dimension of the scalar current first written down by 
Tarrach~\cite{Tarr}. For $n=1$ one reproduces the well-known vanishing of 
the anomalous dimension of the vector current. For $n=3$ one obtains 
agreement with the axial current case derived by Trueman~\cite{True}, and, 
finally, for $n=4$ with the two-loop anomalous dimensions of the 
pseudo-scalar current (with the 't~Hooft-Veltman convention for $\gamma_5$ 
to be discussed later on) which was calculated by Larin~\cite{Lari}. The 
anomalous dimension of the tensor current (with $\Gamma=\sigma_{\mu\nu}$; 
$n=2$) was recently obtained by Broadhurst and Grozin~\cite{BrGr1} in an 
on-shell QCD calculation, who had also checked on the previously published 
results by considering a general Dirac structure for $\Gamma$. Again there 
is agreement.

\subsection{The heavy-light-light irreducible vertex}

Next one needs to calculate the three-quark irreducible proper vertex 
$V_0^{(ir)}$. In this case there are altogether 8 diagrams. Four of them 
are shown in Fig.~2, the other four can be obtained by reflection on the 
heavy quark line which amounts to an exchange of the two light quark legs.

To start with, each quark leg carries a momentum $p_i$ and thus every 
diagram depends on three external momenta. In diagrams (1), (2) and~(4) 
of Fig.~2 we can safely set $p_1=p_2=0$ without introducing any infrared 
singularities, since they are naturally regularized by the off-shell 
energy $\omega$ of the heavy quark. In diagram~(3) we again set $p_2=0$. 
Power counting arguments imply that the contribution of diagram~(3) contains 
a term proportional to $\ln(p_1^2)$. We thus have to keep $p_1$ finite. 
The integrals in diagram~(3) cannot entirely be reduced to two-point 
two-loop integrals, as is the case for diagrams~(1), (2) and~(4). In fact 
the general structure of the two-loop diagram~(3) is given by
\begin{eqnarray}\label{ir3}
{\sl Ir}_3=T_{\mu\nu\rho}\int\frac{dk\,dl}{(2\pi)^{2D}}
  \frac{(l-p_1)^\mu l^\rho}{(l-p_1)^2l^2k^2(\omega+kv)}\cdot
  \frac{(k-l)^\nu}{(k-l)^2},
\end{eqnarray}
where $T_{\mu\nu\rho}$ represents a string of Dirac matrices. The 
simplest way to evaluate this integral is to add and subtract an integral 
which has the same structure as the integral shown in Eq.~(\ref{ir3}) but 
with the simplified propagator $k^\nu/k^2$ instead of $(k-l)^\nu/(k-l)^2$. 
The simplified integral can be easily calculated even for $p_1\ne 0$ 
because it factorizes into two one-loop type integrals. The advantage is 
that the difference of the two integrals with propagators $(k-l)^\nu/(k-l)^2$ 
and $k^\nu/k^2$ is IR-finite as $p_1\rightarrow 0$ and thus one can set 
$p_1=0$ when evaluating the difference integral. The latter integral can 
then be solved by the standard methods. The results of our calculation are 
again presented in coefficient form,
\begin{equation}
V^{(ir)}_0=2\sum_{j=0}^4\left(\sum_{i=1}^4
\lambda_G^2C_B^2b^{(ir)}_{i,j}\right)(s(D-2n))^j,
\end{equation}
where the factor of 2 accounts for the contributions of the remaining four 
reflected diagrams not included in Fig.~2. For the definition of the index~$j$ 
we refer to Eqs.~(\ref{vertll}) and~(\ref{contractions}). The index $i$ 
identifies the relevant Feynman diagram in Fig.~2. The coefficients 
$b^{(ir)}_{i,j}$ are listed in the Appendix~C. For the $Z$-factors and for 
the anomalous dimension we obtain
\begin{eqnarray}\label{anomirr}
Z^{(ir)}_{22}&=&C^2_B(2n^2-8n+9),\\
\gamma^{(2)}_{(ir)}\ =\ -4Z^{(ir)}_{21}&=&-2C^2_B(9n-10+6s)(n-2).\nonumber
\end{eqnarray}
It is interesting to note that the unrenormalized irreducible two-loop 
vertex contains a $1/\epsilon^2$-contribution even though there is no 
corresponding one-loop contribution to the irreducible three-quark vertex. 
The coefficient of the $1/\epsilon^2$-contribution is related to the 
$1/\epsilon$-pole in the total one-loop result by the non-linearity in the 
interplay between Eq.~(\ref{andim12}) and Eq.~(\ref{check}). We have added 
this comment because this interplay of one- and two-loop contributions is a 
feature specific to the baryonic case and does not occur in the mesonic 
case. The result in Eq.~(\ref{anomirr}) completes our calculations.

\section{Anomalous dimensions of $\Lambda_Q$ and $\Sigma_Q$ baryons}

We are now in the position to present our results for the anomalous 
dimension of the baryonic currents in Eqs.~(\ref{currents}). We first sum 
up the results for the light-light, the  heavy-light and the irreducible 
heavy-light-light vertex and obtain 
\begin{eqnarray}\label{gammavres}
\gamma^{(2)}_V(n,s)&=&\frac{C_B}9\Big(
  -9C_B(5n^4-40n^3+106n^2-104n+24+4(n-2)s+32\zeta(2))\nonumber\\&&
  +C_A(18n^4-144n^3+289n^2+128n-516+72\zeta(2))\nonumber\\&&
  +72C_F(n-3)(n-1)+4N_FT_F(n^2-16n+24)\Big).
\end{eqnarray}
To obtain the full two-loop result for the anomalous dimension one has to 
add the anomalous dimensions of the heavy and light quark fields 
from Eq.~(\ref{gammaq}) according to Eq.~(\ref{gammaj}).

Up to this point we have steered clear of the $\gamma_5$-issue because we 
did not want to interrupt the general flow of arguments in the preceding 
sections. When one wants to apply the general result Eq.~(\ref{gammavres}) 
to the $\Lambda_Q$ case in Eq.~(\ref{currents}) one faces the notorious 
problem of how to generalize $\gamma_5$ to $D\ne 4$ dimensions (the 
$\gamma_5$ in the $\Sigma_Q$-type currents is harmless since it is attached 
to the heavy quark line). We shall not commit ourself to any definite 
$\gamma_5$-scheme but give our results for the two generic 
$\gamma_5$-schemes that are being used in the literature, namely the naive 
$\gamma_5$-scheme with an anticommuting $\gamma_5$ (for a general 
exposition of the naive $\gamma_5$-scheme see~\cite{Krei}) and the 
't~Hooft-Veltman-Breitenlohner-Maison (HV) $\gamma_5$-scheme~\cite{tHVt}. 
In the naive $\gamma_5$-scheme one can simply anticommute the $\gamma_5$ in 
the $\Lambda_Q$-currents to the end of the light fermionic string and then 
apply the general result Eq.~(\ref{contractions}). For the total anomalous 
dimension $\gamma_j$ up to two loop order one then obtains (see Table~1 for 
the specific value of the pair $(n,s)$)
\begin{table}
\begin{tabular}{|l|c|c|l|}
\hline
$\Gamma$&$n$&$s$&particles\\\hline\hline
$\gamma^{\rm AC}_5$&$0$&$+1$&$\Lambda_1$\\\hline
$\gamma^{\rm AC}_5\gamma_0$&$1$&$-1$&$\Lambda_2$\\\hline
$\vec\gamma$&$1$&$+1$&$\Sigma_1,\Sigma^*_1$\\\hline
$\gamma_0\vec\gamma$&$2$&$-1$&$\Sigma_2,\Sigma^*_2$\\\hline\hline
$\gamma^{\rm HV}_5$&$4$&$-1$&$\Lambda_1$\\\hline
$\gamma^{\rm HV}_5\gamma_0$&$3$&$+1$&$\Lambda_2$\\\hline
\end{tabular}
\caption{Specific values of the pair $(n,s)$ for particular cases of the
light-side Dirac structure $\Gamma$.}
\end{table}
\begin{eqnarray}
\gamma_{\Lambda1}
  =-8\left(\frac{\alpha_s}{4\pi}\right)+\frac19(16\zeta(2)+40N_F-796)
  \left(\frac{\alpha_s}{4\pi}\right)^2,\label{andimlam1}\\ 
\gamma_{\Lambda2}
  =-4\left(\frac{\alpha_s}{4\pi}\right)+\frac19(16\zeta(2)+20N_F-322)
  \left(\frac{\alpha_s}{4\pi}\right)^2,\label{andimlam2}\\
\gamma_{\Sigma1}
  =-4\left(\frac{\alpha_s}{4\pi}\right)+\frac19(16\zeta(2)+20N_F-290)
  \left(\frac{\alpha_s}{4\pi}\right)^2,\\
\gamma_{\Sigma2}
  =-\frac83\left(\frac{\alpha_s}{4\pi}\right)+\frac1{27}(48\zeta(2)+8N_F+324)
  \left(\frac{\alpha_s}{4\pi}\right)^2. 
\end{eqnarray}
The anomalous dimensions of the spin-$3/2$ currents $\vec J_{\Sigma^*1}$ 
and $\vec J_{\Sigma^*2}$ coincide with those of the spin-$1/2$ currents 
$J_{\Sigma1}$ and $J_{\Sigma2}$, resp., because the respective light-side 
Dirac structures are identical.

In the HV $\gamma_5$-scheme the anomalous dimensions of $J_{\Sigma1}$ 
and $J_{\Sigma2}$ remain unchanged as remarked on before. Because one has
\begin{equation}
\gamma_5=\gamma_0\gamma_1\gamma_2\gamma_3=\frac1{4!}
  \varepsilon_{\mu_1\mu_2\mu_3\mu_4}\gamma^{\mu_1}\gamma^{\mu_2}
  \gamma^{\mu_3}\gamma^{\mu_4}
\end{equation}
in the HV $\gamma_5$-scheme, this particular definition of $\gamma_5$ 
is naturally included in the general representation of the $\gamma$-matrix 
string $\Gamma$ used before in Eq.~(\ref{Dirac}) and Eq.~(\ref{gammavres}). 
In the HV $\gamma_5$-scheme one obtains
\begin{eqnarray}
\gamma_{\Lambda1}
  =-8\left(\frac{\alpha_s}{4\pi}\right)+\frac19(16\zeta(2)-24N_F+260)
  \left(\frac{\alpha_s}{4\pi}\right)^2,\label{andimlam1s}\\
\gamma_{\Lambda2}
  =-4\left(\frac{\alpha_s}{4\pi}\right)+\frac19(16\zeta(2)-12N_F+206)
  \left(\frac{\alpha_s}{4\pi}\right)^2.\label{andimlam2s}
\end{eqnarray}
Comparing Eqs.~(\ref{andimlam1}) and~(\ref{andimlam2}) with 
Eqs.~(\ref{andimlam1s}) and~(\ref{andimlam2s}) one observes that the 
$\zeta(2)$ contributions are not affected by the choice of $\gamma_5$, but 
there are big differences in the remaining contributions between the two 
$\gamma_5$-schemes. Needless to say that physical matrix elements do not 
depend on the choice of $\gamma_5$.

With the availability of the full two-loop result one can check on the 
``quick and easy''-prescription advanced in~\cite{BrGr1} to estimate the 
magnitude of two-loop corrections. This prescription goes by the name of 
``naive non-abelianization''. The idea is to replace $N_F$ by $N_F-33/2$ in 
the easily computed light-quark loop contribution. This prescription is 
based on the observation that the first term of the $\beta$-function is a 
large number and hence the dominant part of any corrections is connected 
with the term proportional to $\beta^{(1)}$. Thus one can hope to arrive at 
a good estimate of a non-abelian result by replacing the abelian 
$\beta^{(1)}$-function by its non-abelian counterpart. We find that this 
prescription gives a good estimate of the full result with an accuracy of 
15--30\% as can easily be checked. An exception is the case $J_{\Sigma2}$ 
(and thereby $\vec J_{\Sigma^*2}$) where the ``naive non-abelianization'' 
estimate is completely off the mark. But note that in the latter case the 
anomalous dimension is an order of magnitude smaller than in the other cases.

As mentioned before, the anomalous dimensions of the baryonic currents 
$J^{\rm HV}$ and $J^{\rm AC}$ are different at two loop order. However, 
the currents can be related to each other by finite 
renormalization~\cite{True,Lari},
\begin{equation}\label{larin}
J^{\rm AC}(\mu)=Z_{\Gamma}J^{\rm HV}(\mu).
\end{equation}
The finite coefficients $Z_\Gamma$ depend on the Dirac structure $\Gamma$, 
which occurs in the currents $J=\bar q\Gamma q$ and $J=(q^{T}C\Gamma q)Q$ 
in the mesonic case and baryonic case, respectively. The coefficients 
$Z_\Gamma$ may be determined by comparing the renormalized matrix elements 
in Eq.~(\ref{larin}) or by calculating the logarithmic derivative of 
Eq.~(\ref{larin}),
\begin{equation}
\frac{d\ln Z_\Gamma}{d\ln\mu}
  =\gamma_{J^{\rm HV}}-\gamma_{J^{\rm AC}}=O_\Gamma.
\end{equation}
Let us introduce a factor $A_{\Gamma}$ through the relation 
$Z_\Gamma=1+A_\Gamma(\alpha_s(\mu)/4\pi)$, where the coefficient $A_\Gamma$ 
can be calculated from 
\begin{equation}\label{agamma}
A_\Gamma\beta(\alpha_s)\left(\frac{\alpha_s(\mu)}{4\pi}\right)=-O_\Gamma,
\end{equation}
where $\beta(\alpha_s)$ was defined in Eq.~(\ref{betadelta}). Using the 
results for the two-loop anomalous dimensions derived before one obtains
\begin{eqnarray}
O_{\gamma_5}
  &=&\frac{16}3C_B(11C_A-4N_FT_F)\left(\frac{\alpha_s}{4\pi}\right)^2,
  \nonumber\\
O_{\gamma_5\gamma_i}=O_{\gamma_5\gamma_0}
  &=&\frac83C_B(11C_A-4N_FT_F)\left(\frac{\alpha_s}{4\pi}\right)^2,\\
O_{\gamma_5\gamma_i\gamma_j}=O_{\gamma_5\gamma_i\gamma_0}&=&0.\nonumber
\end{eqnarray}
For completeness we have also included the Dirac structures 
$\Gamma=\gamma_5\gamma_i$, $\gamma_5\gamma_i\gamma_j$ and 
$\gamma_5\gamma_i\gamma_0$ in our discussion even though they do not 
correspond to the $s$-wave baryon states discussed in this paper. By
calculating $A_\Gamma$ according to Eq.~(\ref{agamma}) one finally obtains
\begin{eqnarray}
Z_{\gamma_5}&=&1-8C_B\left(\frac{\alpha_s}{4\pi}\right),\nonumber\\
Z_{\gamma_5\gamma_i}=Z_{\gamma_5\gamma_0}
  &=&1-4C_B\left(\frac{\alpha_s}{4\pi}\right),\\
Z_{\gamma_5\gamma_i\gamma_j}=Z_{\gamma_5\gamma_i\gamma_0}&=&1.\nonumber
\end{eqnarray}
Note that the finite renormalization coefficients for the baryonic currents 
differ from those in the mesonic case discussed in~\cite{True,Lari} by the 
replacement $C_B\to C_F$.

\section{Conclusion}

We have calculated the anomalous dimensions of static heavy baryon currents 
at two-loop order. The anomalous dimensions are basic ingredients in the 
analysis of QCD sum rules in the heavy baryon sector. In order to improve 
on the accuracy of the analysis of the two-point and three-point sum rules 
in the literature one needs to avail of the next-to-leading order (NLO) 
radiative corrections to these sum rules. Part of the NLO corrections are 
determined by the two-loop anomalous dimensions of the baryon currents 
which have been calculated in this paper. We hope to return to the subject 
of the NLO baryonic sum rule corrections in the near future.

\vspace{.5truecm}

\noindent
{\large \bf Acknowledgments:}
\smallskip\\
We were supported in part by the BMBF, FRG, under contract 06MZ566, and by 
Human Capital and Mobility program under contract CHRX-CT94-0579. We would 
like to thank D.~Broadhurst, A.~Grozin, K.~Chetyrkin, B.~Kniehl, B.~Tausk, 
V.~Smirnov and P.~Gambino for informative discussions.

\newpage

\section*{Appendix A}
\setcounter{equation}{0}
\def\theequation{A\arabic{equation}}
In this appendix we list the coefficient functions of the heavy-light bare 
proper vertex $V_0^{(hl)}$ up to two-loop order. We use the abbreviations 
$E_1=\Gamma(1-\epsilon)\Gamma(1+2\epsilon)$ and 
$E_2=\Gamma(1-\epsilon)^2\Gamma(1+4\epsilon)$, where $\Gamma(x)$ is Euler's 
Gamma-function.
\begin{eqnarray}
b^{(hl)}_0&=&\frac{-2aE_1}{D-4}\\[12pt]
b^{(hl)}_1&=&\frac{4E_1^2}{(D-4)^3}
  -\frac{2(3D^2-24D+44)E_2}{(D-6)(D-4)^3(D-3)}\\[12pt]
b^{(hl)}_2&=&\frac{4E_1^2}{(D-4)^3(D-3)}
  -\frac{4(D-5)(D-2)E_2}{(D-6)(D-4)^3(D-3)}\\[12pt]
b^{(hl)}_3&=&\frac{4E_2}{(D-4)^2(D-3)}\\[12pt]
b^{(hl)}_4&=&\frac{2(D-2)E_2}{D-6)(D-4)^2(D-3)}\\[12pt]
b^{(hl)}_5&=&\frac{-8E_1^2}{(D-4)^2(D-3)}
  +\frac{4E_2}{(D-4)^2(D-3)}\\[12pt]
b^{(hl)}_6&=&\frac{-2(D-2)E_2}{(D-6)(D-4)^2(D-3)}\\[12pt]
b^{(hl)}_7&=&0\\[12pt]
b^{(hl)}_8&=&\frac{E_2}{(D-6)(D-4)^2(D-3)}\\[12pt]
b^{(hl)}_9&=&\frac{-E_2}{(D-6)(D-4)^2(D-3)}\\[12pt]
b^{(hl)}_{10}&=&\frac{-4E_1^2}{(D-4)^2(D-3)}
  -\frac{2DE_2}{(D-6)(D-4)^2(D-3)}\\[12pt]
b^{(hl)}_{11}&=&\frac{-4(D-1)E_2}{(D-6)(D-4)^2(D-3)}
\end{eqnarray}

\section*{Appendix B}
\setcounter{equation}{0}
\def\theequation{B\arabic{equation}}
In this appendix we list the coefficient functions of the light-light bare 
proper vertex $V_0^{(ll)}$ up to two-loop order using the abbreviations 
$Q_1=\Gamma(1-\epsilon)^2\Gamma(1+\epsilon)/\Gamma(1-2\epsilon)$ and 
$Q_2=\Gamma(1-\epsilon)^3\Gamma(1+2\epsilon)/\Gamma(1-3\epsilon)$. 
$\Gamma(x)$ is Euler's Gamma-function.
\begin{eqnarray}
b^{(ll)}_{0,0}&=&(1-a)\frac{(D-2)Q_1}{(D-4)(D-3)}\nonumber\\
b^{(ll)}_{0,1}&=&\frac{-Q_1}{2(D-3)}\\
b^{(ll)}_{0,2}&=&\frac{-Q_1}{2(D-4)(D-3)}\nonumber
  \\[12pt]
b^{(ll)}_{1,1}&=&\frac{4(D-2)Q_1^2}{(D-4)^3(D-3)(D-1)}
  -\frac{4(D-2)(D^3-9D^2+20D-4)Q_2}{(D-6)(D-4)^3(D-3)(D-1)(3D-10)}\nonumber\\
b^{(ll)}_{1,2}&=&\frac{(D^3-12D^2+52D-72)Q_1^2}{2(D-4)^3(D-3)^2(D-1)}
  -\frac{(D^4-13D^3+66D^2-168D+176)Q_2}{(D-6)(D-4)^3(D-3)(D-1)(3D-10)}
  \nonumber\\
b^{(ll)}_{1,3}&=&\frac{Q_1^2}{(D-4)(D-3)^2(D-1)}
  -\frac{2Q_2}{(D-4)(D-1)(3D-10)(3D-8)}\\
b^{(ll)}_{1,4}&=&\frac{Q_1^2}{2(D-4)^2(D-3)^2(D-1)}
  -\frac{Q_2}{(D-4)^2(D-1)(3D-10)(3D-8)}
  \nonumber\\[12pt]
b^{(ll)}_{2,0}&=&-\frac{2(D-2)(D^2-12D+24)Q_1^2}{(D-4)^3(D-3)^2(D-1)}
  \nonumber\\&&\qquad\qquad
  -\frac{8(D-2)(D^3-8D^2+12D+8)Q_2}{(D-6)(D-4)^3(D-3)(D-1)(3D-10)}\nonumber\\
b^{(ll)}_{2,1}&=&-\frac{4(3D^2-18D+26)Q_1^2}{(D-4)^3(D-3)^2(D-1)}
  \nonumber\\&&\qquad\qquad
  +\frac{4(5D^5-71D^4+350D^3-680D^2+296D+320)Q_2}%
    {(D-6)(D-4)^3(D-3)(D-1)(3D-10)(3D-8)}\nonumber\\
b^{(ll)}_{2,2}&=&\frac{(D^2-18D+40)Q_1^2}{(D-4)^3(D-3)^2(D-1)}
  \nonumber\\&&\qquad\qquad
  +\frac{4(4D^4-43D^3+124D^2-40D-160)Q_2}{(D-6)(D-4)^3(D-3)(D-1)(3D-10)(3D-8)}
  \nonumber\\
b^{(ll)}_{2,3}&=&\frac{2Q_1^2}{(D-4)^2(D-3)^2(D-1)}
  -\frac{4DQ_2}{(D-4)^2(D-1)(3D-10)(3D-8)}\\
b^{(ll)}_{2,4}&=&\frac{Q_1^2}{(D-4)^3(D-3)^2(D-1)}
  -\frac{2DQ_2}{(D-4)^3(D-1)(3D-10)(3D-8)}
  \nonumber\\[12pt]
b^{(ll)}_{3,1}&=&\frac{-4(D-2)Q_2}{(D-4)(3D-10)(3D-8)}\nonumber\\
b^{(ll)}_{3,2}&=&\frac{-2(D-2)Q_2}{(D-4)^2(3D-10)(3D-8)}
  \\[12pt]
b^{(ll)}_{4,1}&=&\frac{4(D-2)^2Q_2}{(D-6)(D-4)(D-3)(3D-10)(3D-8)}\nonumber\\
b^{(ll)}_{4,2}&=&\frac{2(D-2)^2Q_2}{(D-6)(D-4)^2(D-3)(3D-10)(3D-8)}
  \\[12pt]
b^{(ll)}_{5,0}&=&-\frac{(D-2)(D^3-16D^2+68D-88)Q_1^2}{(D-4)^3(D-3)^2(D-1)}
  -\frac{2(D-2)(D^2-8)Q_2}{(D-4)^3(D-1)(3D-10)}\nonumber\\
b^{(ll)}_{5,1}&=&\frac{(D-6)(D-2)Q_1^2}{(D-4)(D-3)^2(D-1)}
  +\frac{2(D^2+6D-12)Q_2}{(D-4)(D-1)(3D-10)(3D-8)}\\
b^{(ll)}_{5,2}&=&\frac{(D-6)(D-2)Q_1^2}{(D-4)^3(D-3)(D-1)}
  -\frac{2(D^3-17D^2+62D-56)Q_2}{(D-4)^3(D-1)(3D-10)(3D-8)}
  \nonumber\\[12pt]
b^{(ll)}_{6,0}&=&\frac{2(D-2)^2Q_2}{(D-6)(D-4)(D-3)(3D-10)}\nonumber\\
b^{(ll)}_{6,1}&=&\frac{-4(D-2)^2Q_2}{(D-6)(D-4)(3D-10)(3D-8)}\\
b^{(ll)}_{6,2}&=&\frac{-2(D-2)^2Q_2}{(D-6)(D-4)^2(3D-10)(3D-8)}
  \nonumber\\[12pt]
b^{(ll)}_{7,0}&=&\frac{-8(D-2)^2Q_2}{(D-6)(D-4)^2(D-3)(D-1)(3D-10)}\nonumber\\
b^{(ll)}_{7,1}&=&\frac{16(D-2)^2Q_2}{(D-6)(D-4)(D-3)(D-1)(3D-10)(3D-8)}\\
b^{(ll)}_{7,2}&=&\frac{8(D-2)^2Q_2}{(D-6)(D-4)^2(D-3)(D-1)(3D-10)(3D-8)}
  \nonumber\\[12pt]
b^{(ll)}_{8,0}&=&\frac{(D-2)(7D-6)Q_2}{(D-6)(D-4)^2(D-3)(D-1)(3D-10)}
  \nonumber\\
b^{(ll)}_{8,1}&=&\frac{-2(D-2)(6D-5)Q_2}{(D-6)(D-4)(D-3)(D-1)(3D-10)(3D-8)}\\
b^{(ll)}_{8,2}&=&\frac{-(D-2)(6D-5)Q_2}{(D-6)(D-4)^2(D-3)(D-1)(3D-10)(3D-8)}
  \nonumber\\[12pt]
b^{(ll)}_{9,0}&=&\frac{-(D-2)^2Q_2}{(D-6)(D-4)^2(D-3)(D-1)(3D-10)}\nonumber\\
b^{(ll)}_{9,1}&=&\frac{-2(D-2)Q_2}{(D-6)(D-4)(D-3)(D-1)(3D-10)(3D-8)}\\
b^{(ll)}_{9,2}&=&\frac{-(D-2)Q_2}{(D-6)(D-4)^2(D-3)(D-1)(3D-10)(3D-8)}
  \nonumber\\[12pt]
b^{(ll)}_{10,0}&=&\frac{(D-2)^2(D^2-12D+30)Q_1^2}{(D-4)^3(D-3)^2(D-1)}
  \nonumber\\
  &&\qquad\qquad-\frac{2(D-2)(D^4-17D^3+106D^2-268D+216)Q_2}%
    {(D-6)(D-4)^3(D-3)(D-1)(3D-10)}\nonumber\\
b^{(ll)}_{10,1}&=&\frac{(5D^3-48D^2+138D-116)Q_1^2}{(D-4)^3(D-3)^2(D-1)}\\
  &&\qquad\qquad-\frac{2(11D^5-195D^4+1314D^3-4148D^2+6072D-3264)Q_2}%
    {(D-6)(D-4)^3(D-3)(D-1)(3D-10)(3D-8)}\nonumber\\
b^{(ll)}_{10,2}&=&\frac{2(2D^2-9D+8)Q_1^2}{(D-4)^3(D-3)^2(D-1)}
  -\frac{4(5D^3-34D^2+71D-44)Q_2}{(D-4)^3(D-3)(D-1)(3D-10)(3D-8)}
  \nonumber\\[12pt]
b^{(ll)}_{11,1}&=&\frac{-8(D-2)(D-1)Q_2}{(D-6)(D-4)(D-3)(3D-10)(3D-8)}
  \nonumber\\
b^{(ll)}_{11,2}&=&\frac{-4(D-2)(D-1)Q_2}{(D-6)(D-4)^2(D-3)(3D-10)(3D-8)}
\end{eqnarray}

\section*{Appendix C}
\setcounter{equation}{0}
\def\theequation{C\arabic{equation}}
In this appendix we present the two-loop coefficient functions of the 
irreducible heavy-light-light proper vertex $V_0^{(ir)}$.
\begin{eqnarray}
b^{(ir)}_{1,0}&=&\frac{2E_2}{(D-4)^2}\qquad\qquad b^{(ir)}_{4,j}\ =\ 0
  \\[12pt]
b^{(ir)}_{2,1}&=&\frac{-E_2}{(D-6)(D-4)(D-3)}\qquad
b^{(ir)}_{2,2}\ =\ \frac{-E_2}{(D-6)(D-4)^2(D-3)}\qquad
  \\[12pt]
b^{(ir)}_{3,1}&=&\frac{-(D-4)E_2}{2(D-8)(D-6)(D-3)}
  +\frac{E_1Q_1}{(D-4)(D-3)}\qquad
  \\[12pt]
b^{(ir)}_{3,2}&=&\frac{(D^2-8D+8)E_2}{2(D-8)(D-6)(D-4)^2(D-3)}
  +\frac{E_1Q_1}{(D-4)^2(D-3)}\qquad
\end{eqnarray}

\vspace{1cm}
\centerline{\Large\bf Figure Captions}
\vspace{.5cm}
\newcounter{fig}
\begin{list}{\bf\rm Fig.\ \arabic{fig}:}{\usecounter{fig}
\labelwidth1.6cm\leftmargin2.5cm\labelsep.4cm\itemsep0ex plus.2ex}
\item One-loop (diagram $i=$0) and two-loop (diagrams $i=$1--11) heavy-light 
vertex corrections. Light lines: light quarks; heavy lines: static heavy 
quark; curly lines: gluons; dashed lines: ghost contributions
\item Two-loop heavy-light-light irreducible vertex correction. Light lines: 
light quarks; heavy lines: static heavy quark; curly lines: gluons. The 
remaining four diagrams are obtained by reflection on the heavy quark line
\end{list}


\begin{thebibliography}{99}
\bibitem{David} We would like to thank D.J.~Broadhurst for providing us 
  with a REDUCE implementation of the reccurence-relations given 
  in~\cite{ChTk,BrGr}
\end{thebibliography}
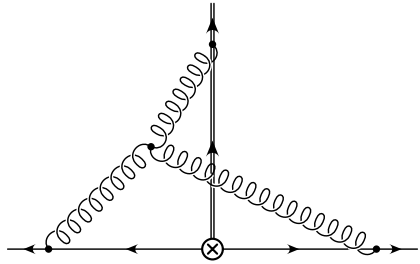
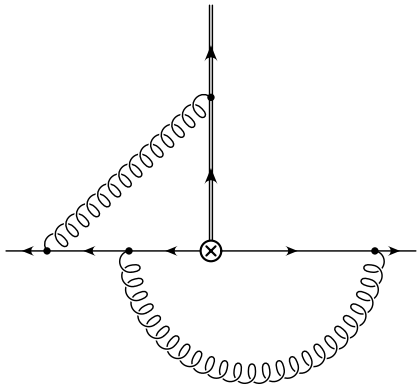
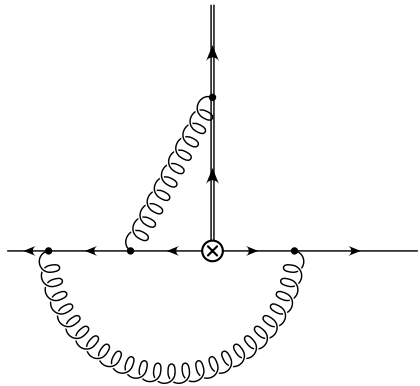
\end{document}